\documentclass[12pt,a4paper,final]{iopart}

\usepackage{iopams}
\usepackage{graphicx}
\usepackage[breaklinks=true,colorlinks=true,linkcolor=blue,urlcolor=blue,citecolor=blue]{hyperref}

\begin{document}

\title{Effect of strain on electronic and thermoelectric properties of few layers to bulk MoS$_{2}$}

\author{Swastibrata Bhattacharyya,Tribhuwan Pandey and Abhishek K. Singh}
\address{Materials Research Centre, Indian Institute of Science, Bangalore 560012, India}

\ead{abhishek@mrc.iisc.ernet.in}

\begin{abstract}

The sensitive dependence of electronic and thermoelectric properties of MoS$_2$ on the applied strain opens up a variety of applications in the emerging area of straintronics. Using first principles based density functional theory calculations, we show that the band gap of few layers of MoS$_2$ can be tuned by applying i) normal compressive (NC), ii) biaxial compressive (BC), and iii) biaxial tensile (BT) strain. A reversible semiconductor to metal transition (S-M transition) is observed under all three types of strain. In the case of NC strain, the threshold strain at which S-M transition occurs increases with increasing number of layers and becomes maximum for the bulk. On the other hand, the threshold strain for S-M transition in both BC and BT strain decreases with the increase in number of layers. The difference in the mechanisms for the S-M transition is explained for different types of applied strain. Furthermore, the effect of strain type and number of layers on the transport properties are also studied using Botzmann transport theory. We optimize the transport properties as a function of number of layers and applied strain. 3L- and 2L-MoS$_2$ emerge as the most efficient thermoelectric material under NC and BT strain, respectively. The calculated thermopower is large and comparable to some of the best thermoelectric materials. A comparison between the feasibility of these three types of strain is also discussed. 

\end{abstract}

%\pacs{71.15.Mb, 72.20.Pa, 71.20.Mq}

\maketitle

\section{Introduction}
Semiconducting 2D materials are very promising for nanoelectronics and nanophotonics applications. Semimetallic graphene, though being the first 2D material to be synthesized and studied \cite {Novoselov2D2005}, cannot be used in these applications as efforts to open the band gap are not yet successful. Due to presence of an intrinsic band gap, molybdenum disulphide (MoS$_2$) a layered material has been explored extensively. The possibility of obtaining free standing single to few layers experimentally \cite{Novoselov2D2005,Frindt1963,frindtMoS1966,Joensen1986,Schumacher1993,ColemanLiquid2011,Ramak2010,Eda2011} with excellent optical absorption and photoconductivity \cite {Frindt1963, korn2011} has attracted recent interest in MoS$_2$. Bulk MoS$_2$ is an indirect band gap semiconductor with a band gap of 1.23 eV \cite{KamBGap1982}. The band gap of MoS$_{2}$ increases with the decrease in number of layers \cite{Han2011} and becomes direct \cite{Splendiani2010}, with a value of 1.9 eV \cite{Mak2010} for monolayer. This semiconducting behaviour has been exploited to various applications such as  field effect transistors with high room-temperature current on/off ratios \cite{RadisavljevicNatnano} and higher on-current density \cite {Liu2011,Yoon2011}, integrated circuits \cite {Radisavljevic2011}, sensors \cite{Late2013,Li2012}, memory cells \cite{Bertolazzi2013} and phototransistors \cite{Yin2012}.

Many electronic and photonics applications require band gap tuning. For MoS$_2$, band gap tuning has been achieved by application of strain \cite{Bhattacharyya2012,Scalise2012} and electric field \cite {Ramasubramaniam2011,Qihang2012}. Semiconductor to metal (S-M) transition was observed in bilayer MoS$_2$ under NC \cite{Bhattacharyya2012} as well as in mono and bilayer MoS$_2$ under BC and BT strains \cite{Scalise2012,Yun2012,Yue2012,Mohammad2013}. Also, for bulk MoS$_2$, application of hydrostatic pressure has shown to reduce the resistivity \cite{Dave2004}. Recently, it has been experimentally shown for the first time that few-layered MoS$_{2}$ undergoes a complete semiconductor to metal transition under the applied hydrostatic pressure \cite{avinash}.

Apart from band gap, strain has also been shown as a potential method to tune the thermoelectric properties of various layered materials \cite{Zahn,Luo,Okuda,MoS2_jap,Guo}. Recent studies have shown that thermoelectric properties of Bi$_2$Te$_3$ \cite{Luo} and Sb$_2$Te$_3$ \cite{Zahn} can further be improved by application of strain. Pardo et al. reported that strain can optimize thermoelectric properties of hole-doped La$_2$NiO$_{4+\delta}$ \cite{Okuda}. MoS$_2$ and other members of dichalogenides family are sought as promising candidates for the next generation thermoelectric devices. They have several advantages, including good environmental compatibility, high thermal as well as chemical stability \cite{mos2_cm} and abundance in nature. They possess high power factors, which are comparable to that of commercially available thermoelectric materials such as Bi$_2$Te$_3$ \cite{Bi2Te3_jmc} and PbTe \cite{MoS2_jap,MoS2_fewL_JCP, mos2_pccp, PbTe_prb}. The theoretical study of Zhang et. al has shown that thermoelectric performance of bulk MoS$_{2}$ can be enhanced by the application of hydrostatic pressure \cite{MoS2_jap}. Recent studies on thermoelectric properties as a function of number of layers for MoS$_{2}$ predicted the possibility of tuning thermoelectric performance by optimizing the layer thickness \cite{MoS2_fewL_JCP,mos2_pccp}. Another important advantage of using few layers is their lower thermal conductivity compared to
bulk \cite{apl_mos2}, which is the most important parameter for efficient thermoelectric performance.

The ability to tune band gap as well as the thermoelectric properties by the application of strain shows great potential of using this 2D material as an electomechanical and thermoelectric device. However, to the best of our knowledge, no studies have investigated the strain induced S-M transition and change in thermoelectric properties of MoS$_{2}$ as a function of applied strain and layer thickness. Here we study the effect of strain and number of layers on electronic as well as thermoelectric properties of MoS$_{2}$. Using first-principles density functional calculations and  Boltzmann transport theory, a comprehensive study is performed to understand the origin of S-M transition under various types of applied strains and its dependence on number of layers. We report that the S-M transition is independent of the number of layers for all the strain types. Although the trend in change in band gap remains same for all the multilayers under the same strain, it is quite distinct for different types of strain. The critical strain for S-M transition increases (decreases) for normal (biaxial) strain. The S-M transition significantly improves the electrical conductivity while keeping a large value of thermopower (250-350 $\mu$V/K), for all the multilayers studied here, giving rise to enhanced thermoelectric transport properties. We optimized the thermoelectric properties with respect to number of layers and strain, and found that 3L- and 2L-MoS$_2$ are the best materials for thermoelectric applications under 0.08 NC and 0.05 BT strain respectively for both \textit{p}- and \textit{n}-type doping.

\section{Method}
The calculations were performed using \textit{ab}-initio density functional theory (DFT) in conjunction with all-electron projector augmented wave potentials \cite{Blochl94,Kresse99}  and the Perdew-Burke-Ernzerhof \cite{Perdew96} generalized gradient approximation to the electronic exchange and correlation, as implemented in the Vienna \textit{Ab initio} Simulation Package (VASP) \cite{Kresse1993}. A well converged Monkhorst-Pack \textbf{k}-point set ($15 \times 15 \times 1$) was used for the Brillouin zone sampling and conjugate gradient scheme was employed to optimize the geometries until the forces on every atom were $\leqslant $0.005 eV/\AA. Sufficient vacuum was used in perpendicular to few-layered MoS$_2$, to avoid spurious interaction among the periodic images. Grimme's DFT-D2 method \cite{Grimme2006} as implemented in VASP was used to incorporate the week van der Waals (vdW) interaction. In Grimme's method, a semi-empirical dispersion potential (D) is added to the conventional Kohn-Sham DFT energy, through a pair-wise force field. 

For transport calculations, Boltzmann transport theory \cite{ashcroft,Ziman} was used, which enables calculation of the temperature and doping level-dependent thermopower and other transport parameters from the electronic structure. All the transport properties were calculated within the constant scattering time approximation (CSTA) \cite{ashcroft,Ziman}. The CSTA is based on the assumption that the scattering time does not vary strongly with energy. It also does not consider any assumptions on temperature and doping level dependence of the scattering time. Within the CSTA, the energy dependence of transport function is described through both the density of states and carrier velocity.
In this theory, the motion of an electron is treated semi-classically, and its group velocity in a specific band is given by
\begin{equation}
 \nu_\alpha(i, \textbf{k}) = \frac{1}{\hbar} \frac{\partial\epsilon(i,\textbf{k})}{\partial \textbf{k}_\alpha}
\label{eq:1}
\end{equation}
where $\epsilon(i,\textbf{k})$ and $\textbf{k}_\alpha$ are the ${i^{th}}$ energy band and $\alpha^{th}$ component of wavevector $\textbf{k}$, respectively. 
From group velocity $\nu_\alpha(i, \textbf{k})$ the thermopower and electrical conductivity can be obtained as
\begin{equation}
S_{\alpha\beta}(T, \mu) = \frac{1}{eT} \frac{\int\nu_\alpha{(i,\textbf{k})}\nu_\beta{(i,\textbf{k})}(\epsilon-\mu)\bigg[-\frac{\partial{f_{\mu}}(T,\epsilon)}{\partial\epsilon}\bigg]d\epsilon}{\int \nu_\alpha{(i,\textbf{k})}\nu_\beta{(i,\textbf{k})}\bigg[-\frac{\partial f_{\mu}(T,\epsilon)}{\partial\epsilon}\bigg]d\epsilon}
\label{eq:2}
\end{equation}

\begin{equation}
 \frac{\sigma_{\alpha\beta}(T, \mu)}{\tau (i, \textbf{k})} = \frac{1}{V} \int e{^2}\nu_\alpha{(i,\textbf{k})}\nu_\beta{(i,\textbf{k})}\bigg[-\frac{\partial{f_{\mu}}(T,\epsilon)}{\partial\epsilon}\bigg]d\epsilon
\label{eq:3}
\end{equation}
where $e$, $T$, $V$, $\tau$ and ${f_{\mu}}$ are the electronic charge, temperature, volume of the unit cell, relaxation time and Fermi-Dirac distribution function, respectively. Brillouin zone was sampled by a well converged 50$\times$50$\times$1 Monkhorst-Pack \textbf{k}-mesh~\cite{pack}. Subsequently, the group velocities were obtained by Fourier interpolation~\cite{fourier} of band energies on converged denser \textbf{k}-grid. These values are used in Eqs.~\ref{eq:2} and~\ref{eq:3}, to calculate the transport properties as implemented in BoltzTraP code~\cite{Madsen}. This approach has been shown to provide a good estimate of thermopower as a function of temperature and carrier concentration in a variety of thermoelectric materials without any adjustable parameters~\cite{Jodin2004,Singh2011,Lee2011,Parker2013,Pandey2013}.

\section{Results and Discussion}
\begin{figure*}[!ht]
\centering
\includegraphics[width=6.5in]{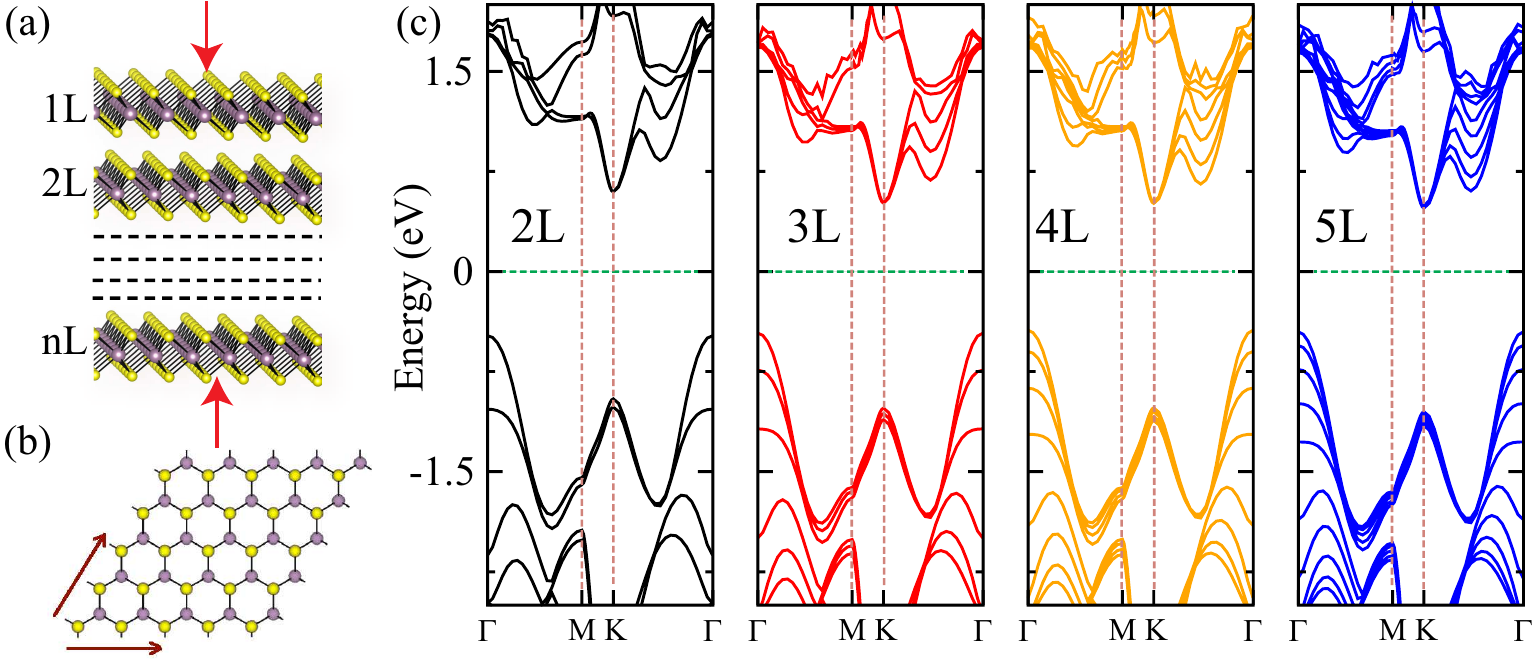}
\caption{ (a) Structure of $n$-layered 2H-MoS$_2$. The Mo and S atoms are shown by purple and yellow spheres. The direction of NC strain is shown by vertical arrows. (b) Top view of MoS$_2$ multilayers with arrows showing the application of biaxial strain. (c) Band structures of $n$L-MoS$_2$ for $n =$ 2, 3, 4 and 5 are shown along the high symmetric K points. The horizontal dotted line represents the Fermi level.}
\label{fig:1}
\end{figure*}

The optimization of the lattice parameters was done using the primitive unit cell of bulk MoS$_2$. The calculated values of $a$ (3.21 \AA) and $c$ (12.42 \AA) are very close to other reported theoretical \cite{Espejo2013} and experimental \cite{Boker2001} results. The multi-layered structures were obtained by arranging MoS$_2$ layers one over another to form AB stacking (Fig. \ref{fig:1} (a)) and named as $n$L-MoS$_2$, where $n$ is the number of MoS$_2$ layers. The NC strain was applied perpendicular to the plane of the multi-layers and along $c$ axis for the bulk case (Fig. \ref{fig:1} (a)). The upper and the lower layers were constrained to move along the normal direction at each NC strain for the multi-layers. For the biaxial cases, equal strain was applied along the $a$ and $b$ axes of the multi-layers as shown in Fig. \ref{fig:1} (b). However, no constraint was used while applying the biaxial strain, allowing the layers to optimize their interlayer distance in each strain. The change in electronic structure and angular momentum resolved density of states (LDOS) under the application of strain was also analyzed for each multilayer MoS$_2$. 

Unstrained MoS$_2$ with more than one layer and bulk has an indirect band gap with valence band maxima (VBM) and conduction band minima (CBM) at $\Gamma$- and K-points of the Brillouin zone, respectively as shown in Fig. \ref{fig:1} (c) for $n$L-MoS$_2$ with $n=$ 2, 3, 4 and 5. The band structure of all these multilayers show that the number of bands forming VBM and CBM increases and are equal to the number of layers. These bands are nearly degenerate at K-point and completely split at the $\Gamma$-point. Furthermore, the spacing between the VBM and VBM$-1$ at the $\Gamma$-point decreases with the increasing number of layers. This nature of the band structure has important implications under the applied strain.

\subsection{Normal compressive strain}

\begin{figure*}[!ht]
\centering 
\includegraphics[width=6.5in]{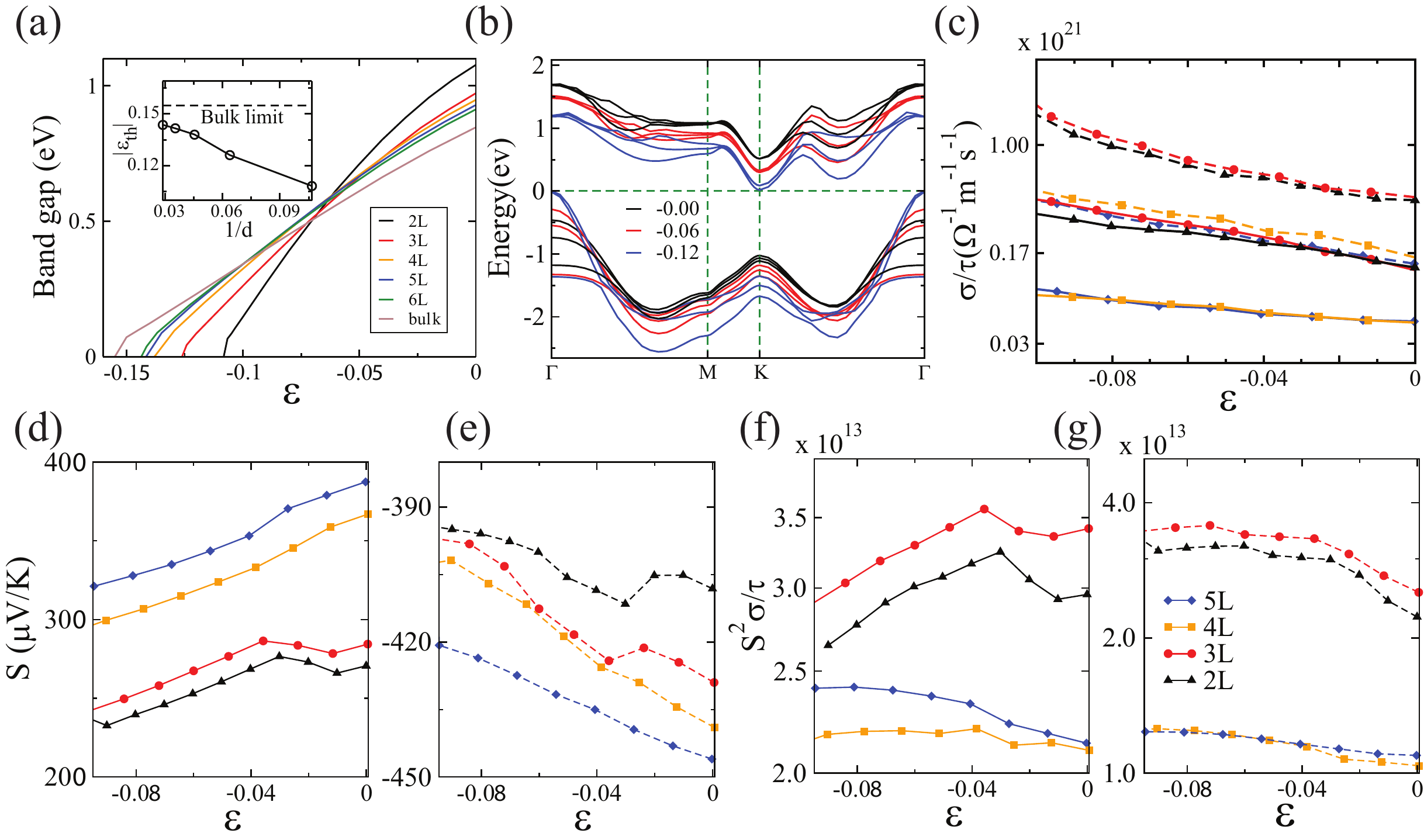}
\caption{(a) Variation of band gap under NC strain is plotted for various multilayers. The variation of $\varepsilon_{th}$ with the inverse of layer thickness ($d$) is shown in the inset. (b) Band structure of 3L-MoS$_{2}$ with increasing NC strain. (c) Calculated $\sigma/\tau$ ratio for MoS$_{2}$ as a function of NC strain for different layer thicknesses. Solid and dotted lines correspond to \textit{p}- and \textit{n}-type of doping, respectively. Thermopower (S) for MoS$_{2}$ as a function of strain at different layer thickness for (d) \textit{p}-type and (e) \textit{n}-type doping. Calculated power factors with respect to relaxation time ($S^{2}\sigma/\tau$) for MoS$_{2}$ as a function of strain for different layer thicknesses for (f) \textit{p}-type and (g) \textit{n}-type doping. In the above transport calculations, the doping level and temperature are fixed at 5 $\times$10$^{19}$ cm$^{-3}$ and 900 K, respectively.}
\label{fig:2}
\end{figure*}

First, we study the effect of NC strain on electronic structure and thermoelectric properties of a few-layer MoS$_2$. The calculated band gap as a function of NC strain is plotted in Fig. \ref{fig:2} (a) for $n$L-MoS$_2$ with increasing number of layers. The applied NC strain leads to S-M transition for all the multilayers and bulk MoS$_2$. The band gap reduces smoothly and becomes zero at a threshold strain $\varepsilon_{th}$, the magnitude of which increases with increasing number of layers and converges towards the bulk limit, i.e., $-0.16$ (Fig. \ref{fig:2} (a) inset). The change in the electronic structure under the application of strain was analyzed for each multilayer and is shown for 3L-MoS$_2$ (which is the best case in terms of thermoelectric performance as shown below) in Fig. \ref{fig:2} (b). With the increase in normal compression, the degenerate bands begin to split (Fig. \ref{fig:2} (b)). The splitting is observed to be more prevalent in the valence band (VB) as compared to the conduction band (CB). The VBM as well as the CBM start to move towards the Fermi level with strain, reducing the band gap smoothly as shown in Fig. \ref{fig:2} (a). An S-M transition occurs when the VBM crosses the Fermi level at the $\Gamma$-point. Upon removal of NC strain, MoS$_{2}$ recovers its semiconducting phase, completely.

The change in electronic structure influences the transport and thermoelectric properties, significantly. Next, we studied the effect of NC strain on transport properties of MoS$_2$ with increase in the number of layers. The calculated $\sigma/\tau$ is shown in Fig. \ref{fig:2} (c). $\sigma/\tau$ increases with increasing NC strain for both \textit{p}- and \textit{n}-type of carrier. Upon application of NC strain both conduction and valence bands become more dispersive giving rise to multi carrier transport, which leads to the increase in conductivity. For the same number of layers, the value of $\sigma/\tau$ was observed to be higher in the case of \textit{n}-type carrier under NC strain. This is expected because conventionally MoS$_2$ is an \textit{n}-type semiconductor. This observation is in good agreement with previous theoretical studies \cite{MoS2_fewL_JCP, mos2_pccp}. The dependence of $\sigma/\tau$ on strain can be explained from the changes in band dispersion as shown in Fig. \ref{fig:2} (b) for 3L-MoS$_2$. Upon application of NC strain, the conduction band gets highly affected in comparison to valence band. CBM becomes more dispersive than VBM increasing the mobility of $n$-type charge carriers, resulting in an enhanced value of conductivity in comparison to the \textit{p}-type carriers. With the increase in strain, 3L- and 2L-MoS$_2$ show highest value of $\sigma/\tau$ (under both \textit{p} and \textit{n}-type doping) due to the contribution from the additional conduction channels, which open up on application of strain. Among all the layers studied here, the factor $\sigma/\tau$ is maximum for 3L-MoS$_2$ over the entire range of NC strains. Assuming relaxation time ($\tau$) to be independent of number of layers, one can conclude that 3L-MoS$_2$ will show the highest electrical conductivity. 

Next, we calculated the thermopower of the unstrained and strained MoS$_{2}$ for different layer thicknesses. The average thermopower ($S$) at 900 K as a function of the NC strain with increasing layer thickness, for \textit{p} and \textit{n}-type doping, are shown in Figs. \ref{fig:2} (d) and (e), respectively. The average thermopower is defined as 1/3 of the trace of the thermopower tensor ($S_{av} = (S_{xx} + S_{yy} + S_{zz})/3$). For these calculations, the doping level is fixed at 5 $\times$ 10$^{19}$ cm$^{-3}$ which is the optimized value for all the layers. The thermoelectric performance peaks at 900 K, and therefore, all the results are presented at this temperature.

\begin{figure*}[!ht]
\centering
\includegraphics[width=6.5in]{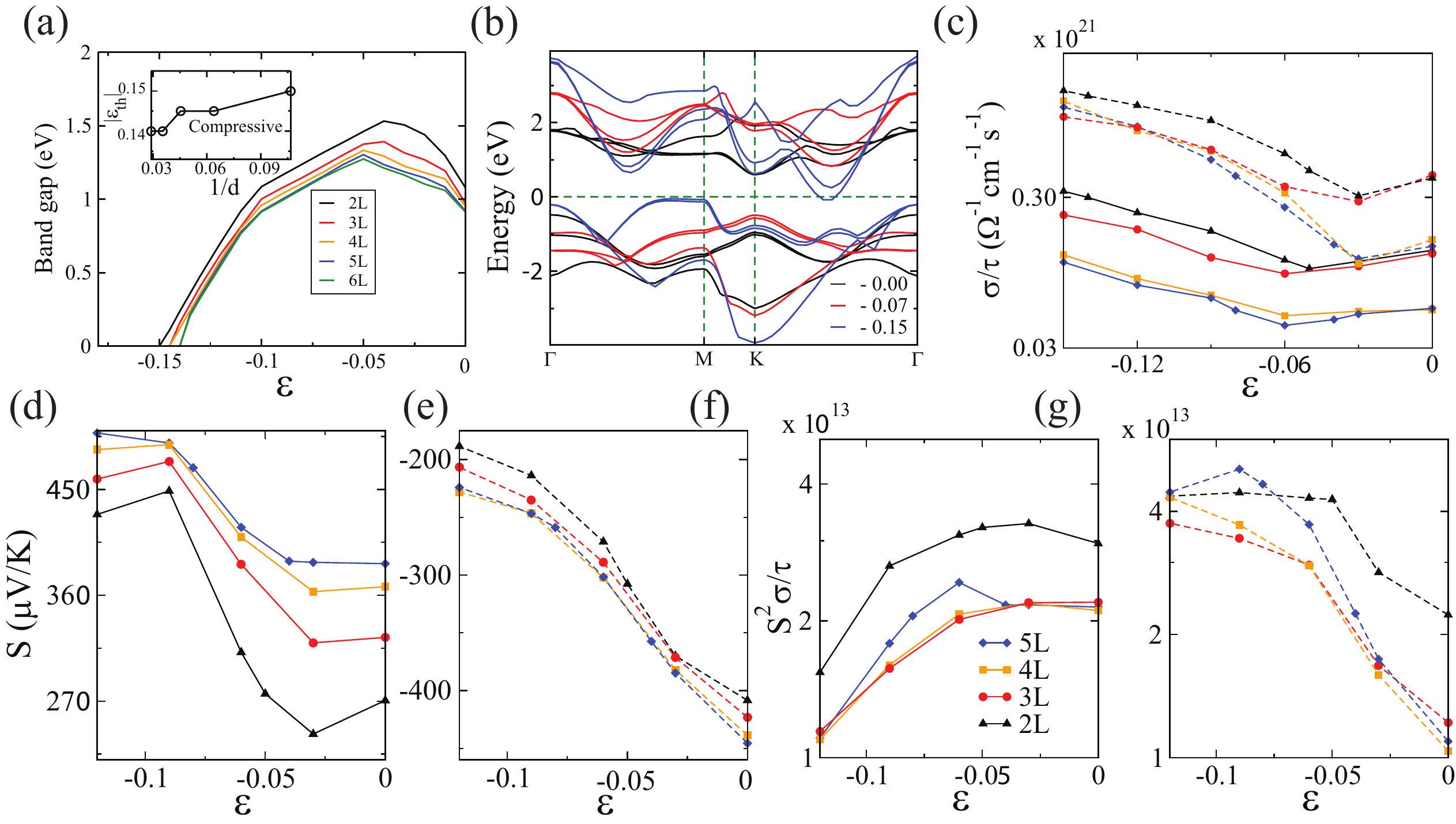}
\caption{(a) Variation of band gap under BC strain is plotted for various multilayers. The variation of $\varepsilon_{th}$ with the inverse of layer thickness ($d$) is shown in the inset. (b) Band structure of 2L-MoS$_{2}$ under increasing BC strain. (c) Calculated $\sigma/\tau$ ratio for BC strain for MoS$_{2}$ as a function of strain for different layer thicknesses. Solid and dotted lines correspond to  \textit{p}- and \textit{n}-type of doping, respectively. Thermopower ($S$) for MoS$_{2}$ as a function of strain at different layer thicknesses for (d) \textit{p}-type and for (e) \textit{n}-type doping. Calculated power factors with respect to relaxation time ($S^{2}\sigma/\tau$) for MoS$_{2}$ as a function of strain for different layer thicknesses for (f) \textit{p}-type and (g) \textit{n}-type doping. In the above transport calculations, the doping level and temperature are fixed at 5 $\times$10$^{19}$ cm$^{-3}$ and 900 K, respectively.}
\label{fig:3}
\end{figure*}
With increase in the number of layers, the thermopower increases due to increase in charge carrier pockets Fig. \ref{fig:2} (b). Under applied NC strain, the changes in the thermopower show the same trend for different layer thicknesses except for 2L- and 3L-MoS$_2$ at lower strains, where the thermopower, first increases and then decreases with peaks at -0.04 and -0.09 for \textit{p}- and \textit{n}-type of carriers, respectively. The absence of such peak for four and higher number of layers indicates the advent of bulk-like behaviour. With the application of strain the band gap starts to decrease, which leads to a slight decrease in thermopower. However, under the strain range studied here, the value of thermopower still remains high, lying in between 250-380 and 390-440 $\mu$V/K for \textit{p}- and \textit{n}-type doping, respectively. Such high values of thermopower are comparable to the best known thermoelectric materials such as PbTe \cite{PbTe_prb} and Bi$_2$Te$_3$ \cite{Bi2Te3_jmc}. These high values of thermopower can be explained by analyzing the band structure, as mentioned previously. Upon application of strain, the bands move close to the Fermi level, giving rise to more electron and hole pockets, which leads to high thermopower.

In order to quantify the effect of strain on thermoelectric performance ($ZT = S^{2}\sigma/\kappa_{total}$, where $\kappa_{total}$ is the total thermal conductivity), we calculate power factor ($S^{2}\sigma$), which is the most dominant term in $ZT$. The power factor represents the ability of a material to produce useful electrical power at a given temperature gradient. The large power factor is indicative of better thermoelectric performance. As a first approximation for power factor ($S^{2}\sigma$), here, the quantity $S^{2}\sigma/\tau$ is calculated at 900K for different layer thicknesses as a function of applied NC strain and is shown in Figs. \ref{fig:2} (f) and (g) for \textit{p} and \textit{n}-type doping, respectively. 3L-MoS$_{2}$ gives the highest value of $S^{2}\sigma/\tau$ due to its highest conductivity and large thermopower for all the NC strain ranges. For 2L- and 3L-MoS$_2$, the power factor peaks at -0.04 and -0.09 strain for \textit{p}- and  \textit{n}-type of doping, respectively. With the applied NC strain, the power factor slightly increases for \textit{p}-type. However, for \textit{n}-type, power factor increases by nearly two times in comparison to unstrained case, attaining a maxima near -0.09 strain. Therefore, NC strain has emerged an effective way to tune both electronic and thermoelectric properties.

\subsection{Biaxial compressive strain}
We further study the effect of BC strain on electronic structure and thermoelectric properties of few layer MoS$_2$.
\begin{figure*}[!ht]
\centering
\includegraphics[width=6.5in]{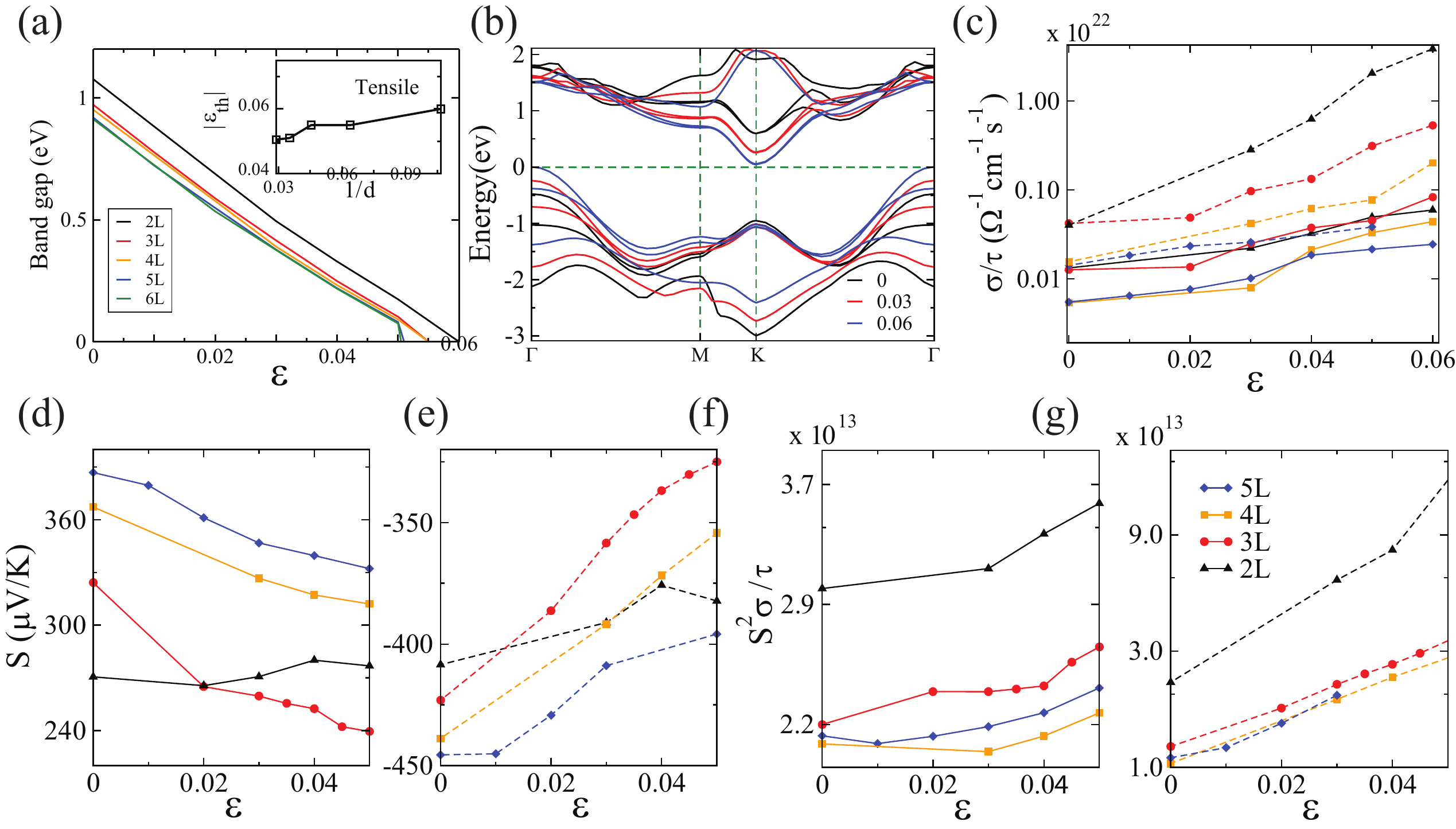}
\caption{(a) Variation of band gap under BT strain is plotted for various multilayers. The variation of $\varepsilon_{th}$ with the inverse of layer thickness ($d$) is shown in the inset. (b) Band structure of 2L-MoS$_{2}$ under increasing BT strain. (c) Calculated $\sigma/\tau$ ratio for BT strain for MoS$_{2}$ as a function of strain for different layer thicknesses. Solid and dotted lines correspond to  \textit{p}- and \textit{n}-type of doping, respectively. Thermopower ($S$) for  MoS$_{2}$ as function of strain at different layer thicknesses for (d) \textit{p}-type and (e) \textit{n}-type doping. Calculated power factors with respect to relaxation time ($S^{2}\sigma/\tau$) for MoS$_{2}$ as a function strain for different layer thicknesses for (f) \textit{p}-type and for (g) \textit{n}-type doping. In the above transport calculations, the doping level and temperature are fixed at 5 $\times$ and 900 K, respectively.}
\label{fig:4}
\end{figure*}
The calculated band gap as a function of BC strain is plotted in Fig. \ref{fig:3} (a) for increasing number of layers. For BC strain, the band gap increases first and then decreases all the way to zero as shown in Fig. \ref{fig:3} (a). The change in band structure is also observed to be very different for BC strain compared to NC strain. The nature of the dispersion changes near the Fermi level with the increase in strain as shown in Fig. \ref{fig:3} (b) for 2L-MoS$_2$. The position of VBM and CBM changes as we move towards higher strain. With the change in the combination of the positions of VBM and CBM, the slope of the band gap vs. strain plot changes as seen in Fig. \ref{fig:3} (a). Five different slopes were seen in the band gap vs. strain plot (Fig. \ref{fig:3} (a)). Initially the position of CBM and VBM are at the high symmetric K- and $\Gamma$-points, respectively, till a strain of -0.02 is reached. During this interval of strain, the band gap increases as shown in Fig. \ref{fig:3} (a) (region of first slope). The position of CBM then changes from K-point to in between K- and $\Gamma$-point, while VBM remains at the $\Gamma$-point, till a strain of -0.04 is reached (region of second slope). Beyond this strain, the band gap begins to decrease as VBM changes to K-point, while CBM remains in between K and $\Gamma$ (region of third slope). At a strain of -0.12 the VBM changes from K- to M-point, while the CBM remains unchanged. At this point, the dispersion of the band becomes flat near the VBM (region of fourth slope). The position of CBM shifts from M- towards $\Gamma$-point, slightly after a strain of -0.13 (region of fifth slope). Finally the CBM crosses the Fermi level at a strain of $\varepsilon_{th}=$-0.15 to give S-M transition. The magnitude of $\varepsilon_{th}$ decreases with increasing number of layers as shown in Fig. \ref{fig:3} (a) inset.
Overall both CBM and VBM move towards the Fermi level. Unlike NC strain, CBM crosses the Fermi level first. The nature of change in the band structure remains the same for all multilayers under BC strain. 

Similar to the band gap, the transport and thermoelectric properties also show different behaviour under BC strain. The plot of $\sigma/\tau$ as a function of BC strain, as shown in Fig. \ref{fig:3} (c) shows opposite trend for both types of charge carriers. $\sigma/\tau$ first decreases with increasing strain and subsequently increases, for both types of carriers. This effect can, once again, be explained on the basis of the band structure. As shown in Fig. \ref{fig:3} (a) the band gap first increases for lower strain values, which leads to initial decrease in conductivity. On further increase in strain, value of $\sigma/\tau$ increases for all the multilayers. The rate of change of conductivity is higher for $n$-type than $p$-type, because the conduction band becomes more dispersive (Fig. \ref{fig:3} (b)) under increasing BC strain giving rise to enhancement in $\sigma/\tau$. However, the bands close to VBM becomes less dispersive (heavy) Fig. \ref{fig:3} (b), which reduces the mobility of holes, and hence, the reduction in $\sigma/\tau$. Under BC strain, 2L-MoS$_{2}$ shows highest value of $\sigma/\tau$ for both $p$ and $n$-type doping.
\begin{figure*}[!ht]
\centering
\includegraphics[width=6.5in]{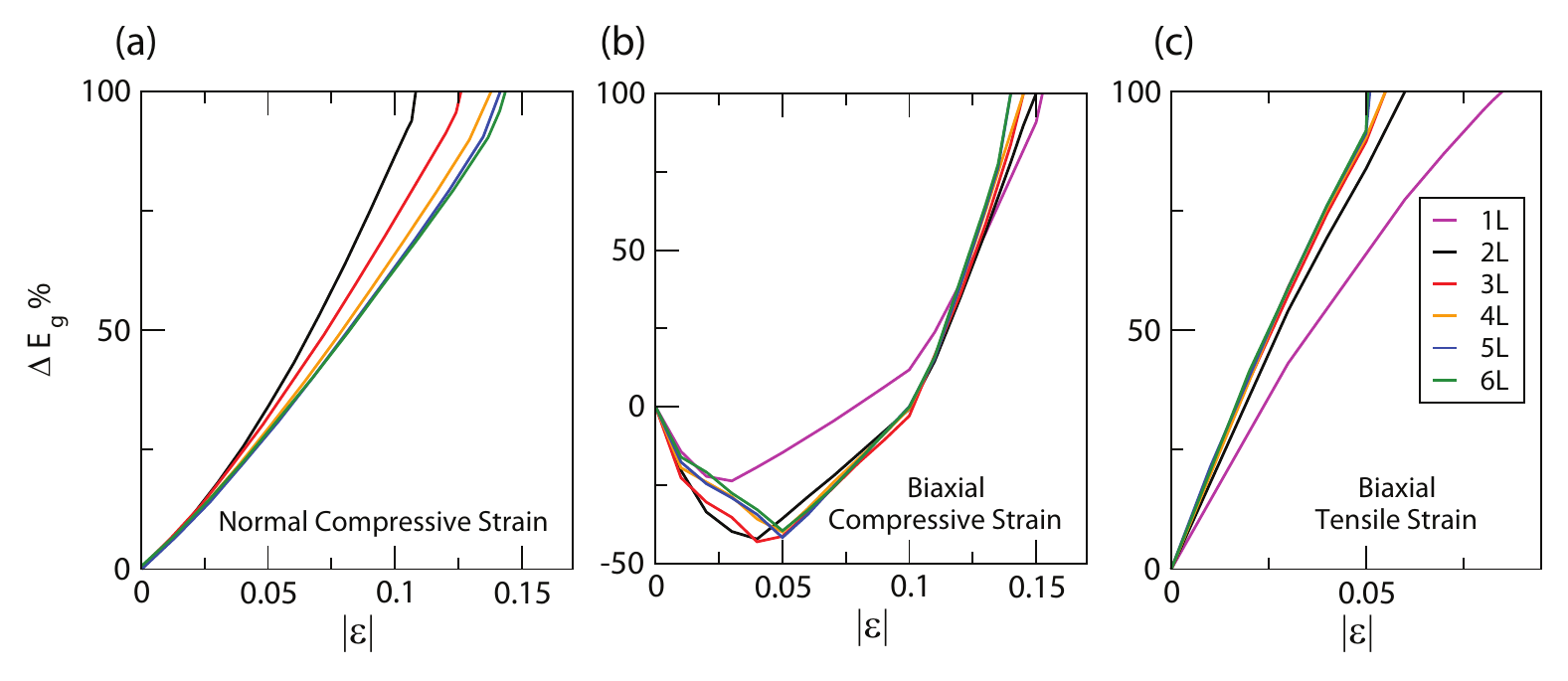}
\caption{ Percentage change in band gap with strain for (a) NC (b) BC, and (c) BT strain types.}
\label{fig:5}
\end{figure*}

Next, we study the effect of BC strain on thermopower as shown in Figs. \ref{fig:3} (d) and (e) for \textit{p}- and \textit{n}-type doping, respectively. The thermopower decreases with increase in strain and saturates at higher strains for $n$-type doping. However, for \textit{p}-type doping, there is an initial decrease with a gradual increase, which finally saturates. This difference in trend of thermopower for \textit{n} and \textit{p}-type of carrier can also be attributed to the band structure, as shown in Fig. \ref{fig:3} (b). Under strain, the conduction band becomes more dispersive. However, some of the valence bands become heavy, for e.g., the band along M-$\Gamma$. The presence of these heavy bands increases the  thermopower with increasing strain for \textit{p}-type doping. A similar trend is also observed for all the multilayers. Furthermore, the power factor increases and attains a maxima around -0.05 and -0.075 BC strain for \textit{p} and \textit{n}-type doping. The highest power factor achieved is for 2L- and 5L-MoS$_2$ under \textit{p}- and \textit{n}-type doping. In particular, under \textit{n}-type doping, the factor $S^{2}\sigma/\tau$ gets enhanced by 4.5 times. Interestingly, these layers also posses highest electrical conductivity, implying that by tuning the electrical conductivity using BC strain, one can enhance the thermoelectric performance, considerably.

\subsection{Biaxial tensile strain}
\begin{figure*}[!ht]
\centering
\includegraphics[width=6.5in]{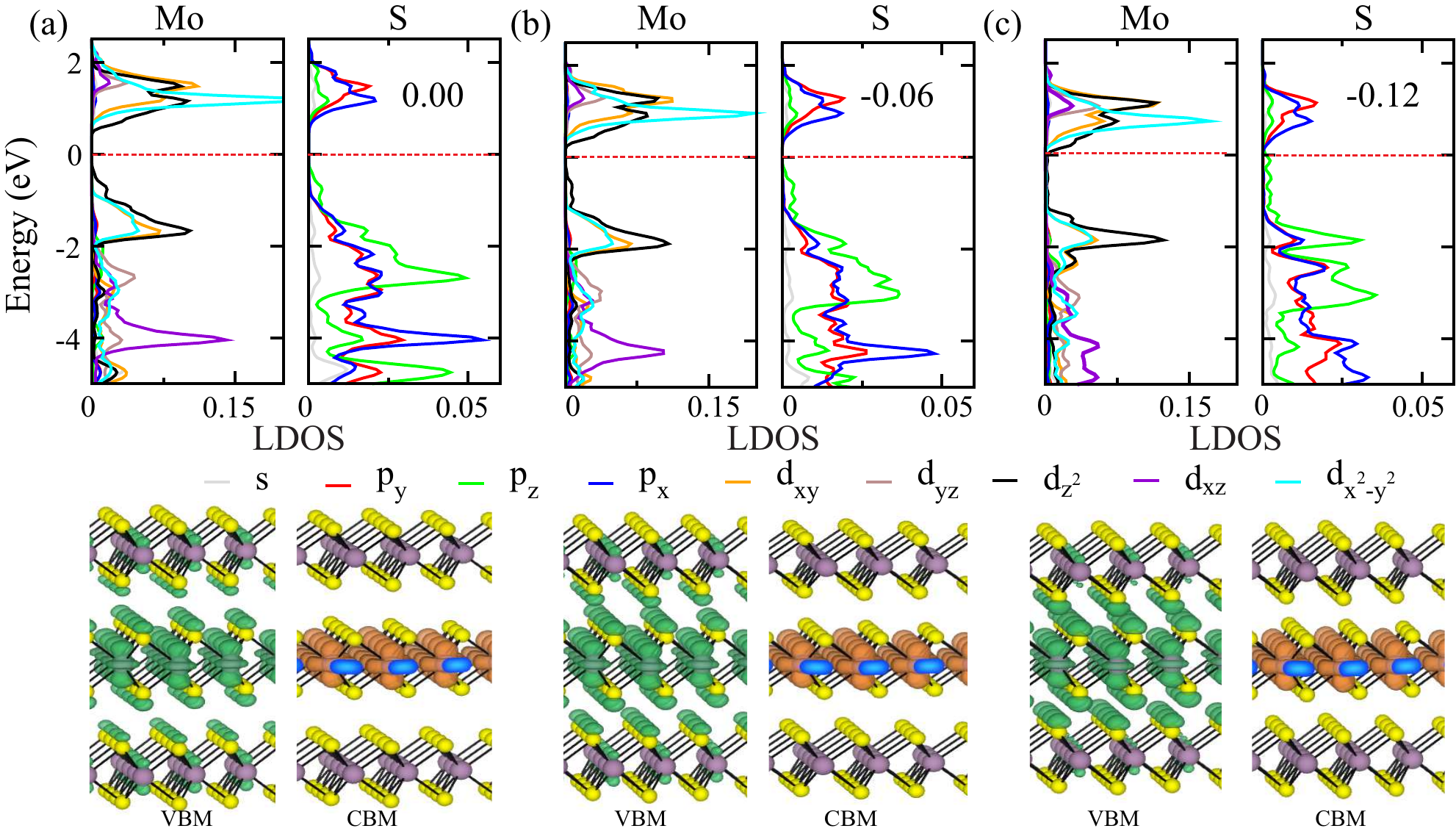}
\caption{Angular momentum projected density of states (LDOS) of Mo and S and corresponding isosurfaces (value of 0.09 eV/\AA$^3$) of the band-decomposed charge density (VBM and CBM) for 3L-MoS$_2$ under NC strains of (a) 0.00 (b) -0.06 and (c) -0.12, respectively. The Fermi level is shown by red dotted lines.}
\label{fig:6}
\end{figure*}
We investigate the effect of BT strain on electronic structure and transport properties of MoS$_{2}$. For all the multilayers, band gap reduces smoothly with the increase in BT strain and becomes zero after a threshold strain $\varepsilon_{th}$, exhibiting S-M transition. Similar to the BC strain, the magnitude of $\varepsilon_{th}$ decreases with increasing number of layers as shown in the inset of Fig. \ref{fig:4} (a). The nature of the dispersion does not change as shown in Fig. \ref{fig:4} (a) for 2L-MoS$_2$. The VBM and CBM remain at K- and $\Gamma$-points, respectively throughout the applied strain range. No significant splitting of bands was observed for this type of strain. The CBM and the VBM move towards the Fermi level as shown in the band structure plots. Similar to the case of NC strain, VBM crosses the Fermi level at $\Gamma$ point at the S-M transition. However, the nature of the change in the band gap vs. strain plot is different in BT strain (Fig. \ref{fig:4} (a)) as compared to the NC and BC strains.

The calculated $\sigma/\tau$ is shown in Fig. \ref{fig:4} (c) for BT strain. $\sigma/\tau$ ratio increases with increasing strain for both types of carriers. This is similar to the NC strain. However, the rate of increase in conductivity is more in the case of \textit{n}-type carriers than the \textit{p}-type. Under this strain, CBM becomes more dispersive than VBM, thereby, increasing the mobility of charge carriers, leading to enhanced value of conductivity in comparison to the \textit{p}-type carriers. More importantly, the factor $\sigma/\tau$ attains a high order of magnitude under BT strain when compared to NC strain indicating better tuning of transport properties. We next calculate the thermopower under BT strain as shown in Figs. \ref{fig:4} (d) and (e) for \textit{p} and \textit{n}-type doping. The thermopower with BT strain shows similar behaviour as that in the case of NC strain. 
For biaxial strain as suggested by high conductivity values, the quantity $S^{2}\sigma/\tau$ becomes maximum for tensile strain. The highest power factor achieved under BT strain is for 2L$-$MoS$_{2}$, which is nearly 1.5 and 4 times higher in comparison with unstrained case for \textit{p}- and \textit{n}-type doping, as shown in Figs. \ref{fig:4} (f) and (g), respectively. The figure of merit $ZT$ closely follows the overall change in the $S^{2}\sigma/\tau$ and hence, it can also be enhanced by the same factor.
\begin{figure*}[!ht]
\centering
\includegraphics[width=6.5in]{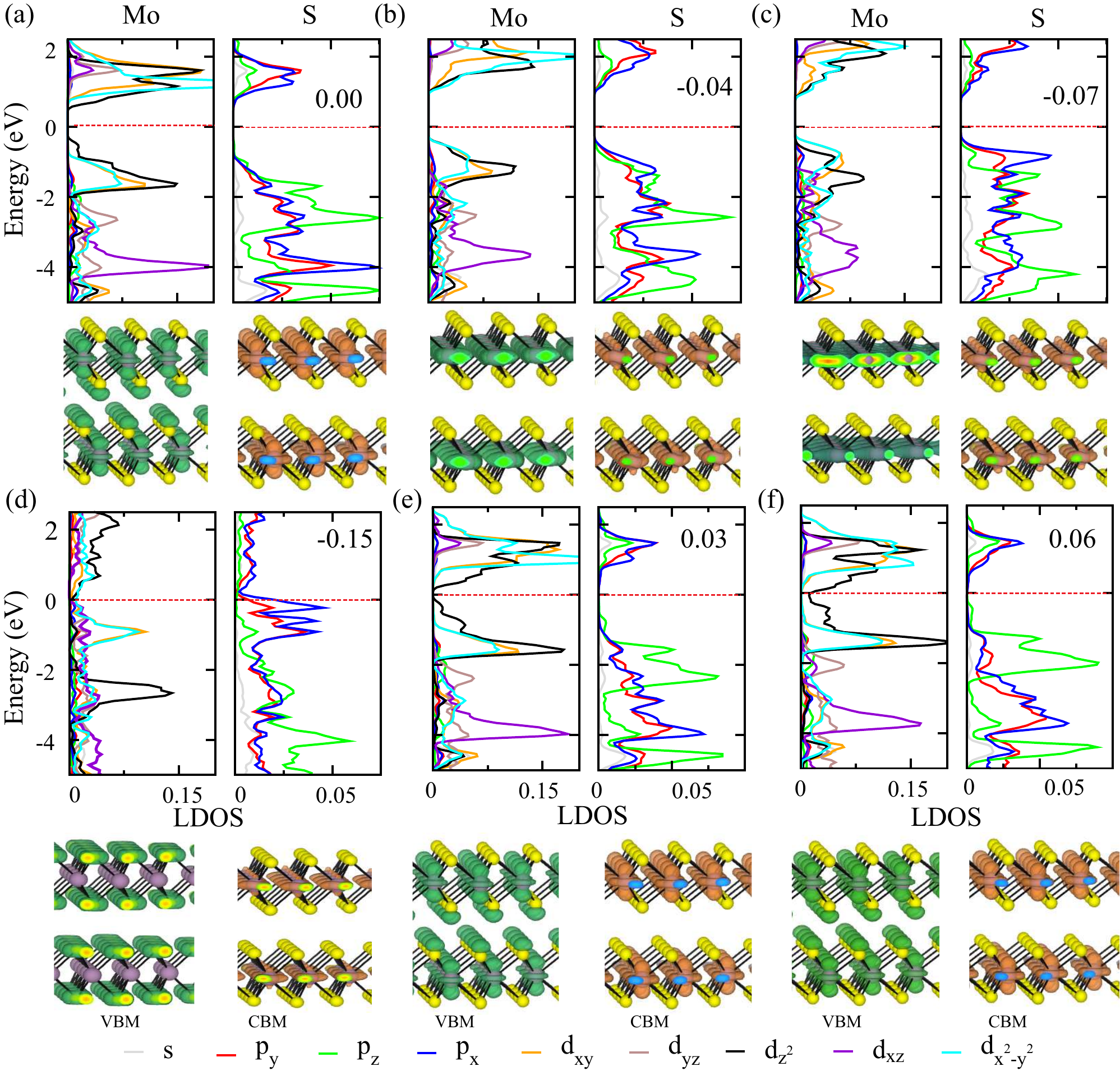}
\caption{ Angular momentum projected density of states (LDOS) of Mo and S and corresponding isosurface (value of 0.09 eV/\AA$^3$) of band-decomposed charge density (VBM and CBM) for 2L-MoS$_2$ under BC and BT strains. (a) For unstrained case. (b)-(d)  Under -0.04, -0.07 and -0.15 BC strains, respectively. (e)-(f) under 0.03 and 0.06 BT strains. The Fermi level is shown by red dotted lines.}
\label{fig:7}
\end{figure*}
\subsection{Mechanism of S-M transition}
In order to compare the sensitivity in terms of change in band gap under the application of NC, BC and BT strains, the percentage reduction in band gap was calculated for all the multilayers considered in this work. The plots of band gap reduction as a function of strain are shown in Figs. \ref{fig:5} (a), (b) and (c) for NC, BC and BT strain, respectively. The curves are almost linear for both the NC as well as the BT strain. The curves are non-linear in case of BC strain, showing a combination of different slopes corresponding to the different regions, as shown in the Fig. \ref{fig:3} (a). From the plots, it is seen that the BT strain leads to the fastest S-M transition as compared to other strains. Increase in the number of layers increases the sensitivity for this kind of strain. 
Though the threshold strain ($\varepsilon_{th}$) is the lowest for BT strain, the application of NC strain is still practically more feasible than the biaxial strains. This is because of the fact that intra-layer covalent bonding between Mo and S is much more stronger than inter-layer vdW bonding. Although the strain required is small in case of BT strain, in comparison to NC strain, more energy and hence pressure will be required to obtain S-M transition as compared to the later case.  

To understand the mechanism for this S-M transition under NC, BC, and BT strain, we analyze the contribution from different molecular orbitals by performing LDOS and band-decomposed charge density calculations. This is shown in Fig. \ref{fig:6} for 3L-MoS$_2$ under NC strain and in Fig. \ref{fig:7} for 2L-MoS$_2$ under biaxial strain. For all unstrained MoS$_2$ multilayers, the Mo-$d$ and S-$p$ orbitals have the highest contribution to the VB as well as the CB as shown in Figs. \ref{fig:6} (a) and \ref{fig:7} (a) for 3L- and 2L-MoS$_2$, respectively. Both VB and CB are mainly composed of Mo-$d_{x^2-y^2}$ and $d_{z^2}$. However, in the case of VB, there is some additional contribution from S-$p_{z}$ orbital. This is clearly seen in the band-decomposed charge density (Fig. \ref{fig:6} (a)), where CBM and VBM mainly originate from around Mo-$d_{x^2-y^2}$ (in-plane lobes) and $d_{z^2}$ (out-of-plane lobes). The contribution of S-$p_{z}$ orbital to the VBM is also indicated by the out-of-plane lobes in the band-decomposed charge density. With the application of NC strain, the LDOS corresponding to VB and CB moves towards the Fermi level resulting in the reduction of band gap. All the orbitals of Mo and S atoms contribute to the shift in CB, while in case of VB the movement towards the Fermi level occurs mainly due to the contribution from Mo-$d_{z^2}$ and S-$p_z$ orbitals. The most dominant interaction happens between the Mo-$d_{z^2}$ orbital of one layer and to the S-$p_z$ orbital of another layer. Similar changes were observed for other layers and bulk MoS$_2$. Based on the LDOS and band-decomposed charge density analysis, one can conclude that the strong inter-layer interaction between Mo-$d_{z^2}$ and S-$p_{z}$ is the main cause for S-M transition under NC strain.

The change in LDOS is completely different in case of BC strain. The tail in LDOS of Mo-$d_{z^2}$, forming the VBM and CBM moves away from the Fermi level as the in-plane interaction increases caused by reduction in Mo-S bond lengths. Due to this movement of Mo-$d_{z^2}$ orbitals, initially the band gap increases for lower strain values. With further increase in strain, LDOS of Mo-$d_{z^2}$ crosses the tail of the LDOS of Mo-$d_{x^2-y^2}$, which now consists both CBM and VBM. The in-plane orbitals begin to interact strongly, causing the shift in the position of VBM from $\Gamma$ to K while the CBM moves from K to in between K and $\Gamma$. This can also be seen in the band-decomposed charge density plots where, the VBM changes to completely planar, while CBM has a slight out-of-plane contribution from Mo-$d_{z^2}$ orbital (Figs. \ref{fig:7} (b) and (c)). With further increase in in-plane interaction, the Mo-$d_{x^2-y^2}$ and S-$p_x$ orbitals hybridize strongly within the layer as seen in Fig. \ref{fig:7} (d). This strong intra-layer hybridization leads to the closing of gap. This phenomenon can be seen in the band-decomposed charge density Fig. \ref{fig:7} (d), where the VBM and CBM show strong in-plane character near S-M transition. 

In the case of BT strain, the VBM and the CBM has maximum contribution from Mo-$d_{z^2}$ orbitals (Fig. \ref{fig:7} (e)). With the increase in BT strain, Mo-$d_{z^2}$ and S-$p_{z}$ orbitals start to hybridize strongly within the same layer. This strong hybridization further leads to S-M transition Fig. \ref{fig:7} (f). Although in the case of both NC and BT strains, the S-M transition is dominated by Mo-$d_{z^2}$ and S-$p_{z}$, but in the case of NC strain the interaction happens between Mo and S atoms from different layers. However, for BT strain this interaction is within the same layer.

\section{Conclusion}
In conclusion, we show the tuning of band gap in multilayer MoS$_2$ under the application of normal as well as biaxial strains. A reversible semiconductor to metal transition is obtained by applying NC, BC and BT strains. The nature of the transition in terms of the electronic structure is different in each kind of strain. For a particular strain, the reduction in the band gap remains unchanged for different number of layers. However, the threshold strain at which the transition occurs, increases (decreases) with increase in number of layers for normal (biaxial) strain.  Threshold strain is the least in case of the BT strain and is the maximum for the BC strain. The mechanism behind the S-M transition is also investigated in terms of angular momentum resolved density of states and band-decomposed charge density.  We demonstrate that inter-layer interaction between Mo-$d_{z^2}$ and S-$p_z$ causes the S-M transition under NC strain. However, S-M transition under BC and BT strain is caused by the strong hybridization of the intra-plane Mo-$d_{x^2-y^2}$ and S-$p_x$ orbitals and Mo-$d_{z^2}$ and S-$p_{z}$ orbitals, respectively. The change in transport properties of multilayers MoS$_{2}$ were also investigated under the application of NC, BC, and BT strain for different layer thicknesses. The strain modifies the dispersion of bands, which improves the thermoelectric performance of the material, significantly. The thermopower results suggest that under NC strain the efficient thermoelectric performance can be achieved at -0.04 and -0.09 \% for \textit{p} and \textit{n}-type doping respectively. The 3L-MoS$_{2}$ shows maximum power factor under NC strain. While for biaxial strain, the enhanced performance could be achieved under 0.05 \% compressive/tensile strain for 2L-MoS$_{2}$. The possibility of tuning of band gap by applying strain provides potential application of MoS$_2$ multilayers not only in thermoelectric but also in sensors and other electromechanical devices. Moreover, the value of the threshold strain provides a means to determine the number of layers in an experimentally grown multilayer MoS$_{2}$. Our results indicate that the maximum thermoelectric performance can be achieved under biaxial strain, which can be achievable in experiments. Recently, it has been shown that biaxial strain can be applied by growing thin films on lattice mismatched substrates \cite{MoS2_apl_strain,local_strain_MOS2,few_layer_mOS2}. Therefore, the complete potential of tuning electronic and thermoelectric properties of this wonderful material, can be realized experimentally.

\section*{Acknowledgements}
This work was supported by the ADA under NPMASS and DST nanomission. The authors thank the Supercomputer Education and Research Centre and Materials Research Centre, IISc, for providing the required computational facilities for the above work.

\section*{References}
\providecommand{\newblock}{}

\end{document}